\documentclass[twocolumn,jgrga]{agutex}
\usepackage[dvips]{epsfig}
\usepackage[mathlines,running]{lineno}
\authorrunninghead{HUSSEIN ET AL.}
\titlerunninghead{Diffusion Coefficients of Energetic Particles}
\begin{document}
%
%
\title{The Influence of Different Turbulence Models on the Diffusion Coefficients of Energetic Particles}
\authors{M. Hussein\altaffilmark{1}, R. C. Tautz\altaffilmark{2}, and A. Shalchi\altaffilmark{1}}
\altaffiltext{1}{Department of Physics and Astronomy, University of Manitoba, Winnipeg, MB, R3T 2N2, Canada}
\altaffiltext{2}{Zentrum f\"ur Astronomie und Astrophysik, Technische Universit\"at Berlin, Hardenbergstra\ss e 36, D-10623 Berlin, Germany}
%
%
%
\begin{abstract}
We explore the influence of turbulence on the transport of energetic particles by using test-particle simulations.
We compute parallel and perpendicular diffusion coefficients for two-component turbulence, isotropic turbulence,
a model based on Goldreich-Sridhar scaling, noisy reduced magnetohydrodynamic turbulence, and a noisy slab model.
We show that diffusion coefficients have a similar rigidity dependence regardless which turbulence model is used and,
thus, we conclude that the influence of turbulence on particle transport is not as strong as originally
thought. Only fundamental quantities such as particle rigidity and the Kubo number are relevant. In the current
paper we also confirm the unified nonlinear transport theory for noisy slab turbulence. To double-check the
validity and accuracy of our numerical results, we use a second test-particle code. We show that both codes provide
very similar results confirming the validity of our conclusions.
\end{abstract}
%
%
%
\begin{article}

\section{Introduction}

There are two fundamental problems in plasma physics and astrophysics, namely the understanding of magnetic turbulence
and the propagation of energetic particles such as cosmic rays through a plasma. It is well-known and accepted that these
two problems are linked to each other. In analytical treatments of particle transport, for instance, a crucial input is
the so-called magnetic correlation tensor describing the turbulence. Energetic particles are assumed to move diffusively
and, thus, a diffusive transport equation is usually used to model their motion (see, e.g., \textit{Schlickeiser} [2002] for a review).
Fundamental quantities in that transport equation are the spatial diffusion coefficients, besides other transport parameters
such as the coefficient of stochastic acceleration. We would like to emphasize that in recent years non-diffusive transport
has been discussed as well (see, e.g., \textit{Zimbardo et al.} [2006], \textit{Pommois et al.} [2007], \textit{Shalchi and Kourakis} [2007],
\textit{Tautz and Shalchi} [2010], \textit{Zimbardo et al.} [2012]).

The turbulent magnetic fields $\delta \vec{B}$, which control the transport of energetic particles, are superposed by a
mean magnetic field $\vec{B}_0$. In the general case one has to distinguish between diffusion along and across that
mean field. In the former case the important quantity is the parallel diffusion coefficient $\kappa_{\parallel}$
and in the latter case the perpendicular diffusion coefficient $\kappa_{\perp}$. Together with the so-called
drift coefficient $\kappa_A$, they form the diffusion tensor.

The knowledge of diffusion parameters is essential for different applications such as the description of diffusive
acceleration at interplanetary shocks (see, e.g., \textit{Zank et al.} [2004], \textit{Dosch and Shalchi} [2010],
\textit{Li et al.} [2012], \textit{Wang et al.} [2012]). In the context of interstellar shocks, such as supernova
remnant shocks, usually simple diffusion models are employed but it was shown recently that the form of the diffusion
coefficient can have a strong impact on the cosmic ray spectrum (see \textit{Ferrand et al.} [2014]). Analytical forms
of the diffusion coefficients are also required for studies of solar modulation and space weather (see, e.g.,
\textit{Alania et al.} [2013], \textit{Engelbrecht and Burger} [2013], \textit{Manuel et al.} [2014],
\textit{Potgieter et al.} [2014], \textit{Zhao et al.} [2014], \textit{Engelbrecht and Burger} [2015]).
To understand the propagation of cosmic rays in different astrophysical environments, the knowledge of transport
parameters is also required (see, e.g., \textit{Shalchi and B\"usching} [2010], \textit{Thornbury and Drury} [2014]).

As mentioned above, a critical quantity entering diffusion theories is the magnetic correlation tensor describing the
turbulence. Therefore, one could conclude that a detailed understanding of turbulence is crucial in the theory of
energetic particles. However, there are also some indications obtained from analytical theory that transport is
universal meaning that the details of turbulence are not important if diffusion coefficients are calculated. Based on
the unified nonlinear transport (UNLT) theory (see \textit{Shalchi} [2010]) it was shown in \textit{Shalchi} [2014] that
diffusion coefficients of magnetic field lines and perpendicular diffusion coefficients of energetic particles only
depend on fundamental length scales of the turbulence and the magnetic field ratio $\delta B / B_0$. This statement
was strengthened by the work of \textit{Shalchi} [2015] where it was shown that the field line diffusion coefficient
is only controlled by the Kubo number and the perpendicular diffusion coefficient of energetic particles is only
controlled by the Kubo number and the parallel mean free path. The details of the spectrum such as spectral shape or
anisotropy are not important as long as the aforementioned scales are finite (see, e.g., \textit{Matthaeus et al.} [2007],
\textit{Shalchi} [2014], and \textit{Shalchi} [2015] for details).

In the current paper we are trying to achieve the following:\\
\begin{enumerate}
\item We perform test-particle simulations for five different turbulence models, namely the two-component model, isotropic
turbulence, Goldreich-Sridhar turbulence, the Noisy Reduced MagnetoHydroDynamic (NRMHD) model, and a noisy slab model.
All models are briefly presented in Section 2 of the current paper. We compute numerically the parallel and perpendicular
diffusion coefficients and compare them with each other. We aim to investigate the influence of the turbulence onto the
different transport parameters.\\
\item In \textit{Shalchi} [2015] a new perpendicular transport regime was discovered for the case of short parallel mean
free paths and small Kubo numbers. This new regime is very similar compared to the scaling originally obtained by
\textit{Rechester and Rosenbluth} [1978]. In the current paper we simulate for the first time particle transport in noisy
slab turbulence and test numerically the aforementioned predictions.\\
\item We employ a second test-particle code, namely the so-called PADIAN code developed by \textit{Tautz} [2010] to compute
the different diffusion coefficients independently. This will allow us to check the validity of our simulations and
to understand how accurate and reliable our findings are.\\
\end{enumerate}

The remainder of this paper is organized as follows.  In Section 2 we discuss the different turbulence models employed
in the current paper. The test-particle code which is used is discussed in Section 3. In Section 4 we show the 
diffusion coefficients along and across the mean magnetic field for the different turbulence models. In Section 5
we use the PADIAN code to check the validity of the results presented in Section 4 and we end with a
short summary and some conclusions in Section 6.
\section{Synthetic Models of Turbulence}
In analytical diffusion theories and also in test-particle simulations, an important quantity is the so-called magnetic
correlation tensor defined as
\begin{linenomath}
\begin{equation}
P_{lm} \left( \vec{k} \right) = \left< \delta B_l \left( \vec{k} \right) \delta B_m^* \left( \vec{k} \right) \right>.
\label{BBcorr}
\end{equation}
\end{linenomath}
Here we have used the ensemble average operator $\langle\dots\rangle$. In the general case, the latter tensor can also depend
on time (see, e.g., \textit{Schlickeiser} [2002] and \textit{Shalchi} [2009] for reviews). In the current paper we use the
magnetostatic approximation. In the theory of particle transport, the latter approximation can be justified by considering
only particles moving much faster than the Alfv\'en speed. Such particle energies and rigidities are considered in the present
paper.

It is still unclear what the exact form of the correlation tensor really is in different physical scenarios. One would
expect that turbulence in the solar system is different compared to turbulence in the interstellar space (see, e.g.,
\textit{Hunana and Zank} [2010] or \textit{Shalchi et al.} [2010] for a more detailed discussion of this matter).
In the following we review five different models which were discussed in the literature before. We do not judge which
of these models is more realistic nor do we claim that our list of turbulence models is complete.
\subsection{The two-component model}
In very early treatments of particle transport, simple models for the turbulence have been employed. \textit{Jokipii} [1966],
for instance, used a so-called {\it slab model} in which the magnetic correlation tensor has by definition the following form
\begin{linenomath}
\begin{equation}
P_{lm}^{slab} (\vec{k}) = g^{slab}(k_{\parallel}) \frac{\delta (k_{\perp})}{k_{\perp}} \delta_{lm},
\label{Plmslab}
\end{equation}
\end{linenomath}
with $l,m=x,y$. Here we have used the {\it Kronecker delta} $\delta_{lm}$ and the {\it Dirac delta} $\delta (k_{\perp})$,
respectively. The other tensor components are zero due to the solenoidal constraint. Furthermore, we have used the
spectrum of the slab modes $g^{slab}(k_{\parallel})$ and we have used the wave vector components along and across
the mean magnetic field $k_{\parallel}$ and $k_{\perp}$, respectively. The slab model is basically a one-dimensional
model in which the turbulent magnetic field depends only on the coordinate along the mean field.

Another model with reduced dimensionality is the so-called {\it two-dimensional model} where we have by definition
\begin{linenomath}
\begin{equation}
P_{lm}^{2D} (\vec{k}) = g^{2D} (k_{\perp}) \frac{\delta (k_{\parallel})}{k_{\perp}} \left( \delta_{lm} - \frac{k_l k_m}{k_{\perp}^2} \right)
\label{Plm2D}
\end{equation}
\end{linenomath}
if $l,m=x,y$ and $P_{lz}=P_{zm}=P_{zz}=0$. In this particular model the magnetic field vector as well as the spatial
dependence are two-dimensional. Above we have used the spectrum of the two-dimensional modes $g^{2D} (k_{\perp})$.

A model which is frequently used to approximate solar wind turbulence is the so-called two-component model in which we have
\begin{linenomath}
\begin{equation}
P_{lm}^{comp} (\vec{k}) = P_{lm}^{slab} (\vec{k}) + P_{lm}^{2D} (\vec{k}).
\label{definecomposite}
\end{equation}
\end{linenomath}
This model is supported by solar wind observations (see, e.g., \textit{Matthaeus et al.} [1990], \textit{Osman and Horbury} [2009a],
\textit{Osman and Horbury} [2009b], \textit{Turner et al.} [2012]), numerical simulations (see \textit{Oughton et al.} [1994],
\textit{Matthaeus et al.} [1996], \textit{Shaikh and Zank} [2007]), as well as analytical work (see \textit{Zank and Matthaeus} [1993]).
To complement the two-component model we have to specify the two spectra $g^{slab}(k_{\parallel})$ and $g^{2D} (k_{\perp})$,
respectively. For the former spectrum we use the form proposed by \textit{Bieber et al.} [1994]
\begin{linenomath}
\begin{equation}
g^{slab}(k_{\parallel}) = \frac{C(s)}{2 \pi} \delta B_{slab}^2 l_{slab} \frac{1}{\left[ 1 + (k_{\parallel} l_{slab})^2 \right]^{s/2}},
\label{slabspec}
\end{equation}
\end{linenomath}
where we have used the slab bendover scale $l_{slab}$, the magnetic field strength of the slab modes $\delta B_{slab}$, and the
inertial range spectral index $s$. The normalization function $C(s)$ is given below. For the spectrum of the two-dimensional
modes we employ the one proposed by \textit{Shalchi and Weinhorst} [2009]
\begin{linenomath}
\begin{equation}
g^{2D}(k_{\perp}) = \frac{2 D(s, q)}{\pi} \delta B_{2D}^2 l_{2D} \frac{(k_{\perp} l_{2D})^{q}}{\left[ 1 + (k_{\perp} l_{2D})^2 \right]^{(s+q)/2}},
\label{2dspec}
\end{equation}
\end{linenomath}
where we have used the two-dimensional bendover scale $l_{2D}$, the magnetic field strength of the two-dimensional modes $\delta B_{2D}$,
and the energy range spectral index $q$. The normalization function is given by
\begin{linenomath}
\begin{equation}
D(s, q) = \frac{\Gamma \left( \frac{s+q}{2} \right)}{2 \Gamma \left( \frac{s-1}{2} \right) \Gamma \left( \frac{q+1}{2} \right)},
\end{equation}
\end{linenomath}
where we have used the Gamma function $\Gamma (z)$. The spectrum (\ref{2dspec}) is correctly normalized if $s>1$ and $q>-1$ are satisfied.
The normalization function of the slab spectrum is a special case of the function $D(s, q)$. The two functions are related to each other
via $C (s) \equiv D(s, q=0)$. To ensure that the ultra-scale and other fundamental turbulence scales are finite, we only consider cases
with $q>1$ (see, e.g., \textit{Matthaeus et al.} [2007] and \textit{Shalchi} [2014]).
\subsection{Isotropic turbulence}
Before a more detailed understanding of turbulence became available, scientists used two simple models, namely the slab model
discussed above and the isotropic model (see, e.g., \textit{Fisk et al.} [1974] and \textit{Bieber et al.} [1988]). In the
latter case the turbulent magnetic field itself depends on all three coordinates of space but there is no preferred direction.
In this case the correlation tensor has the form
\begin{linenomath}
\begin{equation}
P_{lm}^{iso} = \frac{g^{iso}(k)}{8\pi k^2} \left( \delta_{lm} - \frac{k_l k_m}{k^2} \right),
\end{equation}
\end{linenomath}
where we have used the isotropic spectrum $g^{iso}(k)$ which is defined so that
\begin{linenomath}
\begin{equation}
\int_{0}^{\infty} d k \; g^{iso}(k) = \delta B^2.
\end{equation}
\end{linenomath}
For the latter spectrum we employ the form
\begin{linenomath}
\begin{equation}
g^{iso} (k) = 4 D(s, q) l_0 \delta B^2 \frac{\left( k l_0 \right)^q}{\left[ 1 + \left( k l_0 \right)^2 \right]^{(s+q)/2}}
\end{equation}
\end{linenomath}
corresponding to the spectrum used above for two-dimensional turbulence. In the current paper we set $q=3$ if the
isotropic model is considered to ensure again finite turbulence scales. The parameter $l_0$ denotes the bendover
scale.
\subsection{Goldreich-Sridhar turbulence}
The {\it Goldreich-Sridhar model} (see \textit{Goldreich and Sridhar} [1995]) does not predict the exact form of the tensor
of Alfv\'enic perturbations, it only states that most energy will be in the state of {\it critical balance}. Combining the
{\it critical balance} condition with the {\it Kolmogorov cascading}, one gets the relation between $k_{\parallel}$ and $k_{\perp}$,
namely, $|k_{\parallel}| \sim k_{\perp}^{2/3}$.

The caveat here is that the wavenumbers are defined in the global magnetic field frame of reference, this may be misleading.
The critical balance is the condition satisfied only in the frame related to the local magnetic field. In what follows we use
the parallel and perpendicular wavenumbers but understand them in terms of wavelets, which allow the choice of the local magnetic field.
More details concerning this matter can be found in \textit{Lazarian and Vishniac} [1999] as well as in \textit{Kowal and Lazarian} [2010].

\textit{Cho et al.} [2002] have proposed a specific form of the magnetic correlation tensor which is in agreement with the relations
discussed above. \textit{Shalchi} [2013] has suggested to use the following generalization of that form
\begin{linenomath}
\begin{eqnarray} 
P_{lm}^{GS} (\vec{k}) = g^{GS}(k_{\parallel},k_{\perp}) \left( \delta_{lm} - {k_l k_m \over k^2} \right)
\label{smithmatt}
\end{eqnarray}
\end{linenomath}
with
\begin{linenomath}
\begin{eqnarray}
g^{GS}(k_{\parallel},k_{\perp}) & = & \frac{D (s,q)}{2 \pi} l^{3} \delta B^2 \nonumber\\
& \times & \frac{\left( k_{\perp} l \right)^{q-s}}{\left[ 1 + \left( k_{\perp} l \right)^2 \right]^{(s+q)/2}}
e^{-l^{2-s} \left| k_{\parallel} \right| k_{\perp}^{1-s}}
\label{genGS}
\end{eqnarray}
\end{linenomath}
which is only valid for $s=5/3$. Compared to \textit{Cho et al.} [2002], the latter spectrum takes into account a more general
behavior at large scales corresponding to the energy range. This spectrum and similar analytical forms were used before in the theory
of energetic particle transport (see, e.g., \textit{Chandran} [2000], \textit{Cho et al.} [2002], \textit{Shalchi et al.} [2010],
\textit{Sun and Jokipii} [2011], and \textit{Shalchi} [2013]). In \textit{Shalchi} [2014] fundamental length scales of turbulence
were computed for this type of spectrum. There it was shown that the ultra-scale is only finite if $q>1$. Therefore, we set $q=2$
if the Goldreich-Sridhar  model is considered.
\subsection{The NRMHD model}
Above we have discussed the two-dimensional turbulence model. One could argue, that this model is a singular model because there
is no variation of the turbulent field along the z-axis. \textit{Ruffolo and Matthaeus} [2013] proposed a so-called {\it Noisy
Reduced MagnetoHydroDynamic (NRMHD)} turbulence model in which the magnetic correlation tensor (\ref{BBcorr}) has the following form
\begin{linenomath}
\begin{equation}
P_{lm}^{nr} (\vec{k}) = \frac{g^{2D} (k_{\perp})}{2 k_{\perp} K} \Theta \left( K - \left| k_{\parallel} \right| \right) \left( \delta_{lm} - \frac{k_l k_m}{k_{\perp}^2} \right)
\label{Plmfornoisy}
\end{equation}
\end{linenomath}
where we have employed the {\it Heaviside step function} $\Theta (x)$ which is defined so that $\Theta ( x > 0 ) = 1$ and $\Theta ( x < 0 ) = 0$.
Therefore, $\Theta (K - |k_{\parallel}|) = 0$ for $|k_{\parallel}| > K$ in agreement with the model used in \textit{Ruffolo and Matthaeus} [2013].
The function $g^{2D} (k_{\perp})$ is the spectrum of the two-dimensional modes as used above.
It is obvious from the definition (\ref{Plmfornoisy}) that the NRMHD model can be understood as a broadened two-dimensional model.
We can easily recover the pure two-dimensional model discussed above by considering the limit $K \rightarrow 0$.

In the following we use the same spectrum which was used before by \textit{Ruffolo and Matthaeus} [2013] as well as by \textit{Shalchi and Hussein} [2014]
in the context of NRMHD turbulence, namely a spectrum with $q=3$. In this particular case Eq. (\ref{2dspec}) becomes
\begin{linenomath}
\begin{equation}
g^{2D} \left( k_{\perp} \right) = \frac{4}{9 \pi} l_{\perp} \delta B^2
\frac{\left( k_{\perp} l_{\perp} \right)^3}{\left[ 1 + \left( k_{\perp} l_{\perp} \right)^2 \right]^{7/3}}.
\label{spec2}
\end{equation}
\end{linenomath}
The model described here was already used in transport theory to compute field line diffusion coefficients (see \textit{Ruffolo and Matthaeus}
[2013] and \textit{Snodin et al.} [2013]) and perpendicular diffusion coefficients of energetic particle (see \textit{Shalchi and Hussein}
[2014]). An interesting aspect of the NRMHD model is the fact that it contains two characteristic length scales, namely the perpendicular
scale $l_{\perp}$ and the parallel scale $l_{\parallel} = K^{-1}$. It was shown in \textit{Shalchi and Hussein} [2014] and \textit{Shalchi} [2015]
that the scale ratio $l_{\parallel}/l_{\perp}$ has a strong influence on the perpendicular diffusion coefficient.
\subsection{A noisy slab turbulence}
Above we have used the NRMHD model which can be understood as a broadened two-dimensional model. We can combine the same idea with
the slab model which is done in the current paragraph. We define the noisy slab model via (see \textit{Shalchi} [2015])
\begin{linenomath}
\begin{equation}
P_{lm}^{ns} (\vec{k}) = \frac{2 l_{\perp}}{k_{\perp}} g^{slab} (k_{\parallel}) \Theta \left( 1 - k_{\perp} l_{\perp} \right)
\left( \delta_{lm} - \frac{k_l k_m}{k_{\perp}^2} \right)
\label{noisyslab}
\end{equation}
\end{linenomath}
where $l_{\perp}$ is a characteristic length scale for the decorrelation across the mean magnetic field. In the case that one or two
indexes are equal to $z$, the element of the correlation tensor is assumed to be zero as in the pure slab model defined above. If there
is no broadening, corresponding to the case $l_{\perp} \rightarrow \infty$, the noisy slab model corresponds to the usual slab model.
In Eq. (\ref{noisyslab}) we have used again the {\it Heaviside step function} and $g^{slab}(k_{\parallel})$ is the usual spectrum of
the slab modes as it was used above. To study particle diffusion in noisy slab turbulence is interesting because \textit{Shalchi} [2015]
predicted that the ratio $\lambda_{\perp} / \lambda_{\parallel}$ is much smaller in this case compared to other turbulence models.
\section{Test-Particle Simulations}
In the current article, we perform simulations to obtain parallel and perpendicular diffusion coefficients numerically. In computer
simulations three steps have to be performed in order to obtain diffusion parameters, which are
\begin{enumerate}
\item A specific turbulence model has to be simulated by employing the approach described below. Here, we consider the slab/2D composite
model, isotropic turbulence, Goldreich-Sridhar turbulence, the NRMHD model, and a noisy slab model.
\item The Newton-Lorentz equation has to be solved numerically for an ensemble of particles to obtain their orbits.
\item From these test-particle trajectories, one can obtain the diffusion coefficients in the different directions of space.
\end{enumerate}
This method of generating the turbulence and simulating the motion of test-particles was used before (see, e.g., 
\textit{Giacalone and Jokipii} [1994], \textit{Micha\l ek and Ostrowski} [1996], \textit{Reville et al.} [2008],
\textit{Tautz} [2010]) and is different compared to the {\it grid method} used by other authors (see, e.g,
\textit{Qin et al.} [2002a] and \textit{Qin et al.} [2002b]).
\subsection{General remarks}
In order to calculate the turbulent magnetic field at the position of the charged particle $\vec{x}$, one can use the Fourier representation
\begin{linenomath}
\begin{equation}
\delta \vec{B} \left( \vec{x} \right) = \int d^3k \; \delta \vec{B} \bigl( \vec{k} \bigr) e^{i \vec{k} \cdot \vec{x}}.
\end{equation}
\end{linenomath}
In order to benefit from symmetry, it is preferred to evaluate this integral either in spherical or cylindrical coordinates. Since we are dealing
with a numerical treatment, integrals are replaced by sums. The basic idea is to generate random magnetic fluctuations by superposing a large number
of plane waves with different and random polarizations  and phases. The dimensionality of the used turbulence model matters since it determines to
how many sums the integral will break up. For example, in turbulence models with reduced dimensionality, such as slab or two-dimensional models,
and for isotropic turbulence, the integral can be replaced by a single sum because only one independent wave vector component controls the turbulent
magnetic field. On the other hand, Goldrich-Sridhar, NRMHD, and noisy slab models are more complicated. This is due to the fact that two wave vector
components are relevant, namely $k_{\parallel}$ and $k_{\perp}$. Therefore, an extra sum is required making the simulations more time consuming.
In the following we elaborate on the technical details for the different turbulence models.
\subsection{Slab, two-dimensional, and isotropic turbulence}
In order to simulate turbulence models with reduced dimensionality, we use the method described in \textit{Hussein and Shalchi} [2014a] and
\textit{Hussein and Shalchi} [2014b]. More details about the numerical approach used in such simulations can be found in \textit{Tautz and Dosch} [2013].
In numerical treatments of the transport, it is convenient to use dimensionless quantities instead of the physical quantities used above.
If the slab model is considered, for instance, physical quantities are the parallel component of the wave vector $k_{\parallel}$ and the parallel
particle position $z$. In the test-particle code those quantities are replaced by $k_{\parallel} \rightarrow k_{\parallel} l_{slab}$ and
$z \rightarrow z/l_{slab}$, respectively. More details concerning other quantities such as the magnetic rigidity are discussed below.

In numerical work the turbulent magnetic field vector is calculated via
\begin{linenomath}
\begin{equation}
\delta \vec{B} \left( \vec{x} \right) = \sqrt{2} \delta B \sum_{n=1}^{N} A(k_n) \hat{\xi}_n \cos \left[ \vec{k}_n \cdot \vec{x} + \beta_n \right].
\label{turb}
\end{equation}
\end{linenomath}
The parameter $N$ corresponds to the number of simulated wave modes and the quantity $A(k_n)$ denotes the amplitude function (see below).
In Eq. (\ref{turb}) we have also used the wave vector $\vec{k}_n = k_n \hat{k}_n$ with the random wave unit vector
\begin{linenomath}
\begin{eqnarray}
\hat{k}_n=
\left(
\begin{array}{c}
\sqrt{1-\eta_n^2}\cos \phi_n \\
\sqrt{1-\eta_n^2}\sin \phi_n \\
\eta_n
\end{array}
\right).
\end{eqnarray}
\end{linenomath}
Furthermore, we have used the random phase $\beta_n$ and the polarization vector
\begin{linenomath}
\begin{eqnarray}
\hat{\xi}_n=
\left(
\begin{array}{c}
-\sin \phi_n \cos\alpha_n + \eta_n \cos\phi_n \sin\alpha_n \\
\cos \phi_n \cos \alpha_n + \eta_n \sin\phi_n \sin\alpha_n \\
-\sqrt{1-\eta_n^2} \sin \alpha_n
\end{array}
\right)
\label{polarization} 
\end{eqnarray}  
\end{linenomath}
with $\eta_n=\cos\theta_n$. The angles $\theta_n$, $\phi_n$, and $\alpha_n$ can have a specific value or they are random angles depending on the
simulated turbulence model (see Table \ref{anglestab} of the current paper for the used values). 

The amplitude function $A(k_n)$ used above depends on the spectrum $G (k_n)$ via 
\begin{linenomath}
\begin{equation}
A^2(k_n)=G(k_n)\Delta k_n \left( \sum_{\mu=1}^{N} G(k_{\mu}) \Delta k_{\mu} \right)^{-1}.
\end{equation}
\end{linenomath}
For the spectrum we use a form corresponding to the analytical models described above, namely
\begin{linenomath}
\begin{equation}
G(k_n)=\frac{k_n^q}{(1+k_n^2)^{(s+q)/2}}.
\end{equation}
\end{linenomath}
The parameters $q$ and $s$ are energy and inertial range spectral indexes as described above. For the slab modes we use $q=0$, for two-dimensional
modes $q=2$, and for isotropic turbulence $q=3$. A \textit{Kolmogorov} [1941] spectrum with $s=5/3$ is used in all cases. To simulate slab
turbulence we set the polar angle $\theta_n=0$ corresponding to $\eta_n = 1$. For two-dimensional turbulence, we set $\eta_n=0$
and $\alpha_n=0$. For isotropic turbulence all angles are randomly generated. The used values and the meaning of $k_n$ in
the different models are summarized in Table \ref{anglestab}. The composite model is created upon superposing slab and two-dimensional
modes via Eq. (\ref{definecomposite}). In the current paper we set $\delta B_{slab}^2 / B_0^2 = 0.2$ and $\delta B_{2D}^2 / B_0^2 = 0.8$
as suggested by \textit{Bieber et al.} [1996].

Eq. (\ref{turb}) is used to compute the magnetic field at the considered position. An alternative approach to compute turbulent magnetic
fields is based on a {\it Fast Fourier Transform} to replace the summation (see, e.g., \textit{Decker and Vlahos} [1986], \textit{Decker} [1993]).

\subsection{Anisotropic three-dimensional models}
In the current paper we also simulate anisotropic three-dimensional models such as the Goldreich-Sridhar model, NRMHD turbulence,
and the noisy slab model. In such cases we replace the single sum in Eq. (\ref{turb}) by a double sum. Now the turbulent magnetic
field vector is given by
\begin{linenomath}
\begin{eqnarray}
\delta \vec{B} \left( \vec{x} \right) & = & \sqrt{2} \delta B \sum_{m=1}^{M}\sum_{n=1}^{N} A(k_n,k_m) \hat{\xi}_n \nonumber\\
& \times & \cos \left( k_n \cos\phi_n x + k_n \sin\phi_n y + k_m z + \beta_{n,m} \right)
\label{turbulence}
\end{eqnarray}
\end{linenomath}
corresponding to cylindrical coordinates where $k_n$ represents the perpendicular wave number $k_{\perp}$ and $k_m$ represents
the parallel wave number $k_{\parallel}$. In Eq. (\ref{turbulence}) we have used again the polarization vector $\hat{\xi}_n$ given
by Eq. (\ref{polarization}) but set $\eta_n=0$ therein. The parameters $M$ and $N$ denote the number of wave modes in parallel
and perpendicular directions, respectively. In the case of NRMHD and noisy slab turbulence, we set $\alpha_n=0$ to ensure that
$\delta B_z=0$. For Goldrich-Sridhar turbulence, however, $\alpha_n$ is a random angle.

The function $A(k_n,k_m)$ used in Eq. (\ref{turbulence}) represents the wave amplitude and parameter $\beta_{n,m}$ denotes the
random phase as before. For the amplitude function $A(k_n,k_m)$ we employ
\begin{linenomath}
\begin{equation}
A^2(k_n,k_m)=\frac{G(k_n) k_n \Delta k_m \Delta k_n} {\sum_{\mu=1}^{M} \sum_{\nu=1}^{N} G(k_{\nu}) k_{\nu} \Delta k_{\mu} \Delta k_{\nu}}
\label{amp}
\end{equation}
\end{linenomath}
and the turbulence spectrum $G(k_n)$ is in the case of the NRMHD model given by
\begin{linenomath}
\begin{equation}
G(k_n)=\frac{k_n^{q-1}}{\left( 1 + k_n^2 \right)^{(s+q)/2}}.
\label{spectrum}
\end{equation}
\end{linenomath}
In the current paper we set $q=3$ as originally used in \textit{Ruffolo and Matthaeus} [2013]. Furthermore, we cut off the spectrum
in the parallel direction by using a maximum wave number corresponding to the parameter $K$ used in Eq. (\ref{Plmfornoisy}).

For the turbulence model based on Goldrich-Sridhar scaling we employ the spectrum
\begin{linenomath}
\begin{equation}
G(k_n,k_m)=\frac{k_n^{q-s}}{\left( 1 + k_n^2 \right)^{(s+q)/2}} e^{-|k_m| k_n^{1-s}}
\end{equation}
\end{linenomath}
with $s=5/3$ and $q=2$ as discussed above.

For the noisy slab model, we use the same spectrum used for slab turbulence, but we cut off the spectrum in the perpendicular
direction by using a maximum wave number corresponding to the parameter $l_{\parallel}/l_{\perp}$.
\subsection{Further parameters and accuracy issues}
For all anisotropic three-dimensional models, $\Delta k_m$ and $\Delta k_n$ are the spacings between wave numbers.
In our simulations we use a logarithmic spacing in $k_m$ and $k_n$ so that
\begin{linenomath}
\begin{equation}
\frac{\Delta k_n}{k_n}= \exp\Bigg[{\frac{\ln(k_{n,max}/k_{n,min})}{N-1}}\Bigg]
\label{k_step}
\end{equation}
\end{linenomath}
and the same for $k_m$. It is important that parallel wave numbers are distributed fine enough so that the so-called {\it resonance condition}
is satisfied. The resonance condition occurs in quasilinear treatments of the transport and states that parallel scattering occurs only
if $\mu R_L k_{\parallel} = 1$. In the latter condition we have used the unperturbed Larmour radius $R_L$ at $\mu = 0$ and the pitch-angle
cosine $\mu$. In the simulations we have to ensure that a large amount of wave numbers are close to the corresponding $k_{\parallel}$.

The size of the box is restricted by the so-called {\it scaling condition} that ensures that no particles travel beyond the maximum size
of the system, $L_{max}=k_{min}^{-1}$. This is ensured via the relation $\Omega t_{max} k_{min} R_L<1$, which corresponds to $vt_{max} < L_{max}$.
In parallel and perpendicular directions we used $k_{min}=10^{-5}$, leading to a relatively large box to ensure that finite
box size effects do not occur. For the maximal wavenumber we used $k_{max} = 10^3$ in both directions.

The simulations contain further parameters controlling the accuracy. We have to specify the number of wave modes in the parallel
direction $N$ and perpendicular direction $M$, respectively. For the simulations performed for slab, two-dimensional, and
isotropic turbulence, there is only one maximum wave number. In this case we have used $N=512$ which is enough to
satisfy the aforementioned conditions. For the anisotropic three-dimensional models, the double sum makes it more challenging
concerning computational time. For the NRMHD model, we have used $N=256$ and $M=32$, respectively. We have performed test
runs with $M$ up to $128$ and no significant differences were noticed. For the simulations performed for Goldrich-Sridhar
and nosiy slab turbulence we have used $N=32$ and $M=256$.

The procedure described so far can be used to compute the turbulent magnetic field at the position of the charged particle.
To obtain the trajectories of the energetic and electrically charged particles, we solve the Newton-Lorentz equation numerically
for $1000$ particles. The latter equation has the form
\begin{linenomath}
\begin{equation}
\frac{d \vec{p}}{d t} = \frac{q}{c} \vec{v} \times \vec{B}[\vec{x}(t)]
\label{newtonlor}
\end{equation}
\end{linenomath}
where $\vec{x(t)}$ is the position of the particle in Cartesian coordinates. In Eq. (\ref{newtonlor}) we replace the turbulent magnetic
field by using either Eq. (\ref{turb}) or (\ref{turbulence}). It is worth noting that we are using the relativistic version of the
Lorentz force and so the momentum is given by $\vec{p}=\gamma m\vec{v}$. The simulations are done using a Monte Carlo code where each
particle is given a random initial position, pitch angle cosine $\mu$, and turbulence angles. After injection, particles are traced for
a sufficiently long time, around tens of thousand of gyro-periods for the particle to overcome the ballistic regime and move diffusively.
For our numerical integrator we use Runge-Kutta of 4th order which keeps truncation error relatively small and under control. 

We have traced 1000 particles for a maximum running time of $\tau_{max} = \Omega t_{max} = n \times 10^{4}$ with integer $n \in {1,2,...,10}$
depending on the particle rigidity. When the particle is less energetic, i.e. have lower rigidity, it takes more time to move diffusively
especially in the perpendicular direction. After that we have averaged over all the 1000 realizations we have been using. To estimate the error
calculated for the diffusion coefficients, we follow the method explained in \textit{Tautz} [2010]. This method takes into account the averaging
procedure over turbulence manifestation and over all particles. Within our calculations the error was relatively small given the fairly large
number of turbulence modes integrated over and the number of realizations used. This allowed for diffusion to be fairly consistent and stable
so deviations from the mean diffusion coefficient where hardly noticed. For that reason we don't include errors when plotting out final graphs.
\section{Diffusion Coefficients obtained from Test-Particle Simulations}
In the current section we perform the test-particle simulations described in Section 3. In order to distinguish these results from the
simulations presented in Section 5, we refer to the code used here as the simulations by Hussein \& Shalchi. In the following we
compute the parallel mean free path $\lambda_{\parallel}$ which is related to the parallel spatial diffusion coefficient $\kappa_{\parallel}$
via $\lambda_{\parallel} = 3 \kappa_{\parallel} / v$, as well as the perpendicular mean free path $\lambda_{\perp}$ and the ratio
$\lambda_{\perp} / \lambda_{\parallel}$. In Tables \ref{MHcomposite}-\ref{MHnoisyslab} we summarize our numerical results. For all simulations
we set $\delta B = B_0$ and $s = 5/3$. The values for the energy range spectral index $q$ are shown in the corresponding table. Some models
contain two scales $l_{\parallel}$ and $l_{\perp}$. The used values for the ratio $l_{\perp} / l_{\parallel}$ are also listed in the
corresponding table.

\begin{figure}
\centering 
\includegraphics[width=0.48\textwidth]{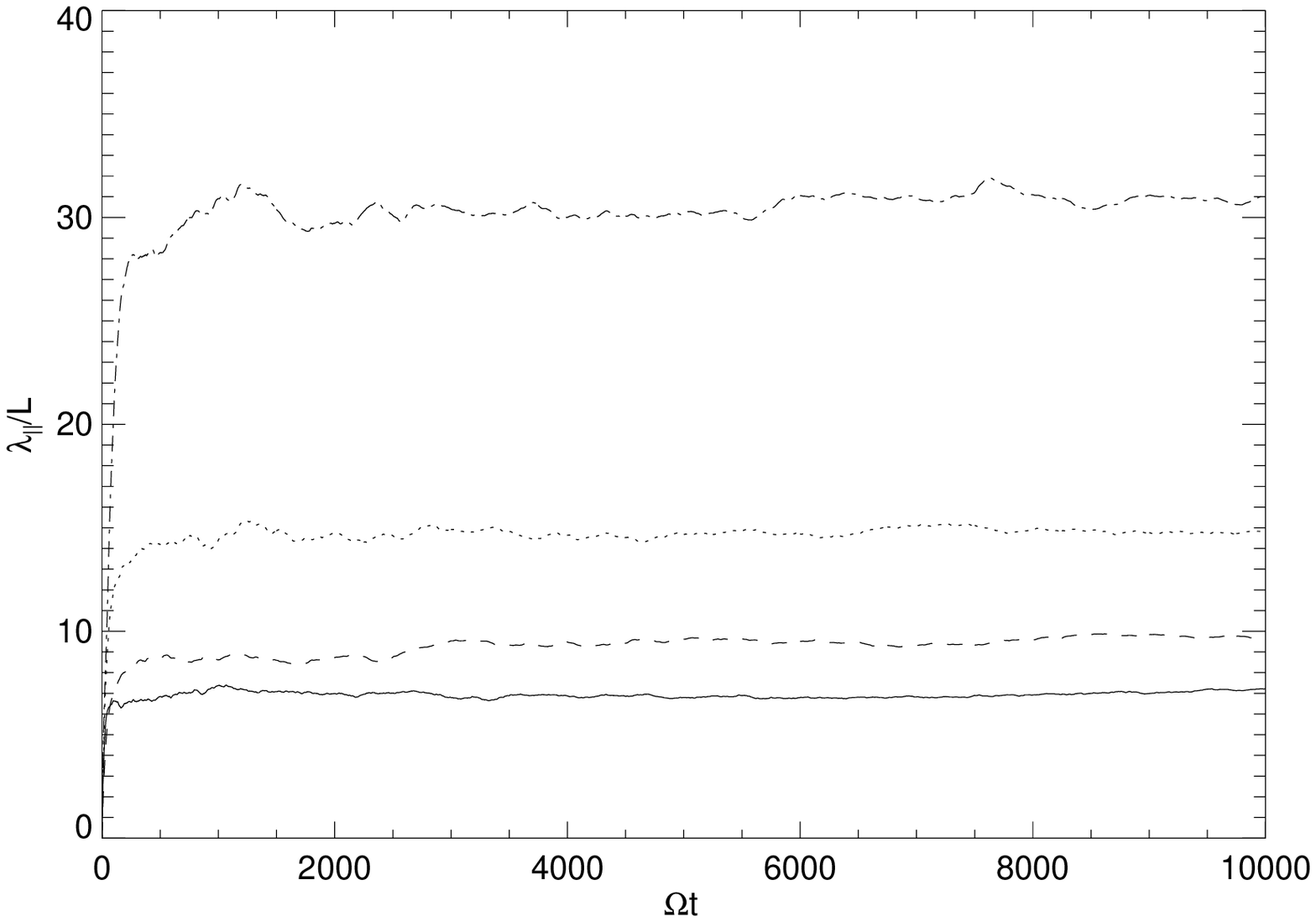}
\caption{The running parallel mean free paths from the Hussein \& Shalchi simulations. Shown are the results obtained for
slab/2D turbulence (dotted line), isotropic turbulence (solid line), Goldreich-Sridhar turbulence (dashed line),
and NRMHD turbulence (dash-dotted line).}
\label{Lparavstime}
\end{figure}

\begin{figure}
\centering 
\includegraphics[width=0.48\textwidth]{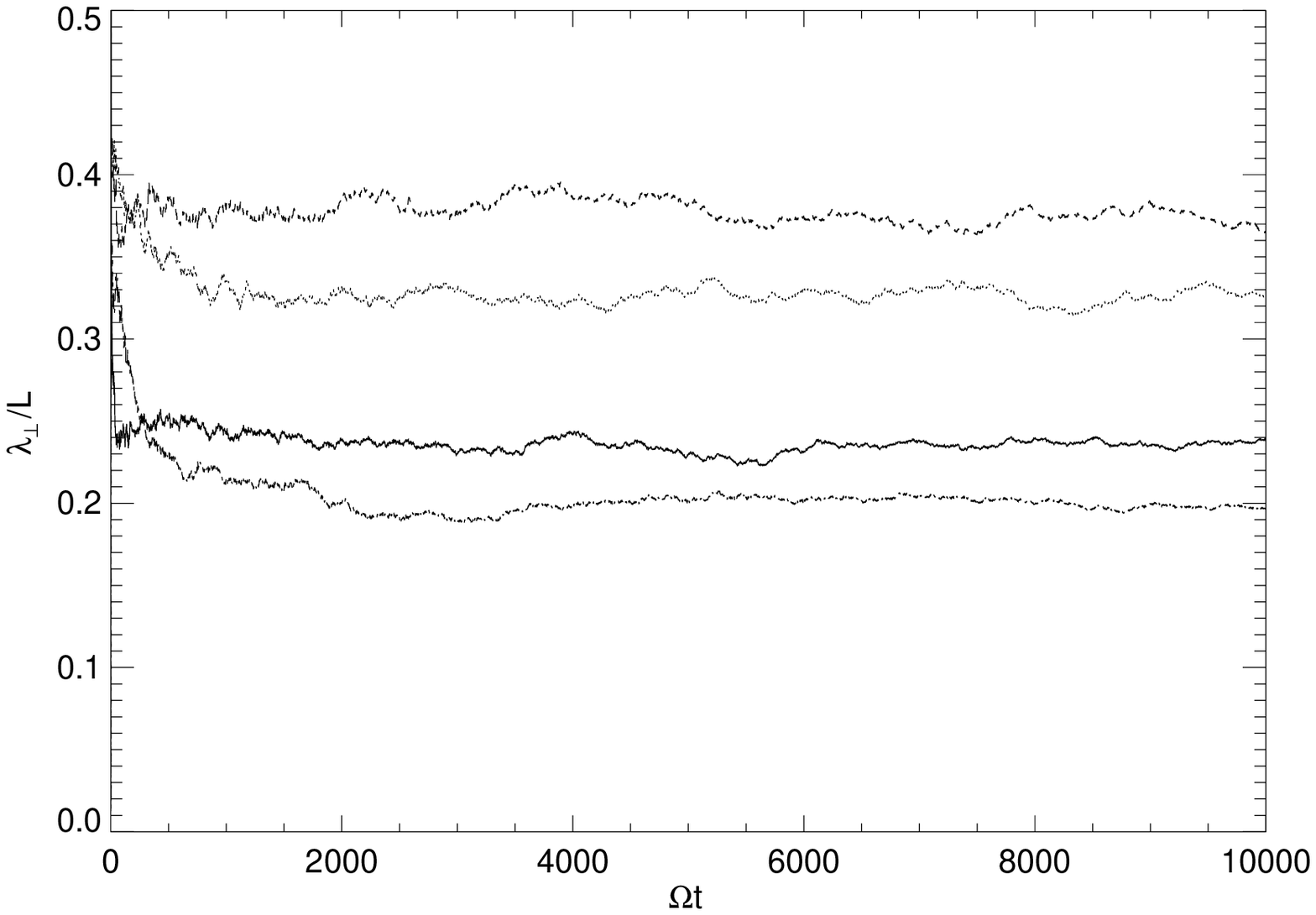}
\caption{The running perpendicular mean free paths from the Hussein \& Shalchi simulations. Shown are the results obtained for
slab/2D turbulence (dotted line), isotropic turbulence (solid line), Goldreich-Sridhar turbulence (dashed line), and
NRMHD turbulence (dash-dotted line).}
\label{Lperpvstime}
\end{figure}

The main aim of the current section is to explore the rigidity dependence of the parallel and perpendicular mean free path,
respectively. However, to ensure that we indeed obtain diffusive transport, we also show the (running) diffusion coefficients
versus time. In Fig. \ref{Lparavstime} we have shown the parallel diffusion coefficient versus the dimensionless time $\Omega t$
for $R=1$. Clearly the diffusion parameters become constant in time after the well-known initial ballistic regime.
Therefore, we conclude that transport is indeed diffusive for the considered cases. Fig. \ref{Lperpvstime} shows the
perpendicular diffusion coefficient versus time. Again we find diffusive transport after the ballistic regime.

\begin{figure}
\centering 
\includegraphics[width=0.48\textwidth]{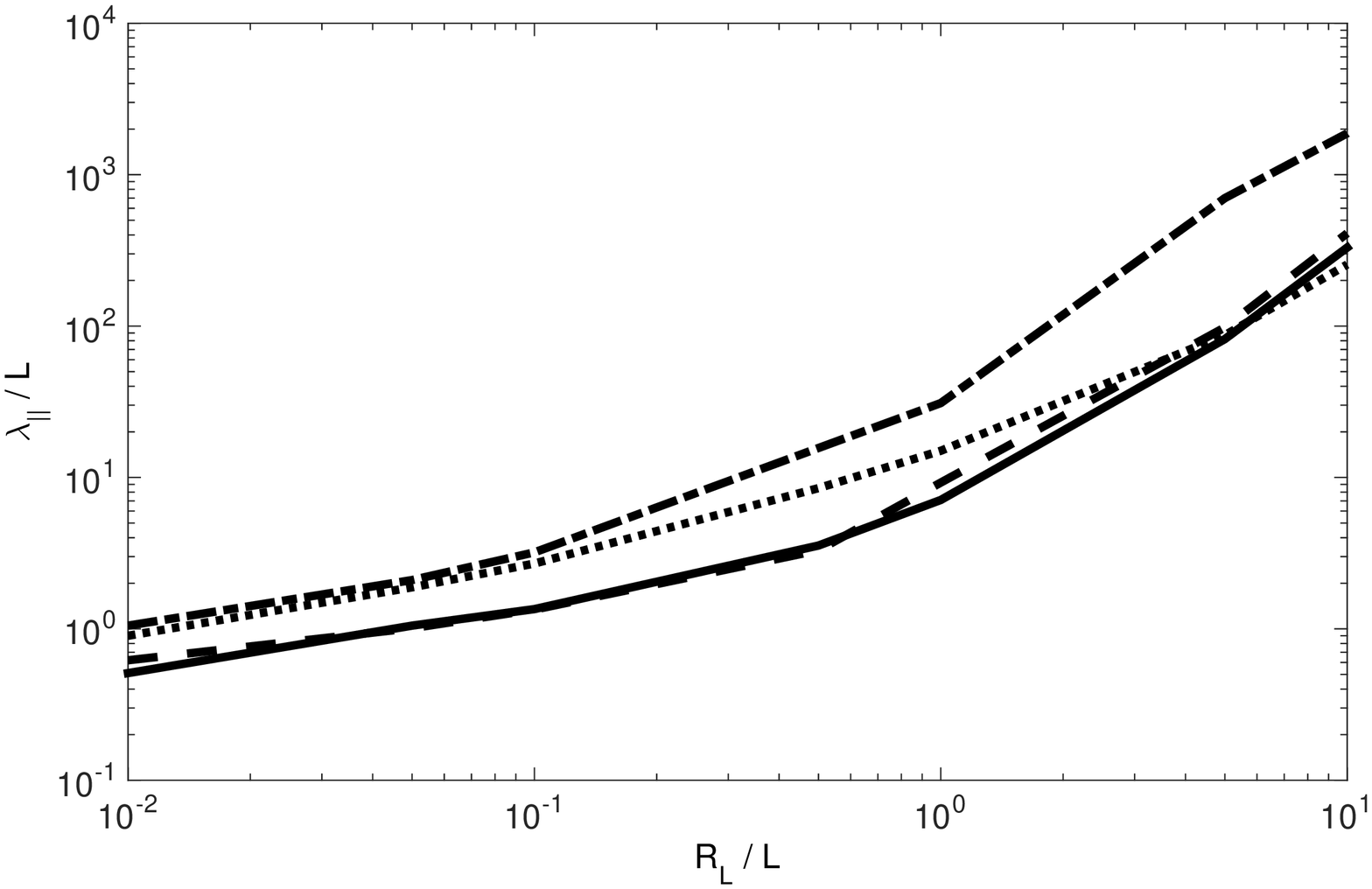}
\caption{The parallel mean free paths from the Hussein \& Shalchi simulations. Shown are the results obtained for
slab/2D turbulence (dotted line), isotropic turbulence (solid line), Goldreich-Sridhar turbulence (dashed line),
and NRMHD turbulence (dash-dotted line).}
\label{LparaMH}
\end{figure}

In Fig. \ref{LparaMH} we show the parallel mean free paths for the different turbulence models except the noisy slab model which is discussed
separately (see below). All quantities are normalized with respect to $L$ which stands for the corresponding turbulence scale (e.g., $L=l_0$
for isotropic turbulence). In all cases the parallel diffusion coefficient increases with increasing rigidity. The rigidity dependence is
approximately the same in each case and only the absolute values are slightly different. The only exception is the parallel mean free path
obtained for the NRMHD model which is clearly larger at high rigidities. We would like to point out that in this particular turbulence model,
there is no turbulence for certain parallel wave numbers. Therefore, there is often no gyro-resonant interaction between the energetic particles
and magnetic fields. This could explain the differences between the parallel diffusion coefficients shown in Fig. \ref{LparaMH}. The analytical
investigation of parallel diffusion in NRMHD turbulence will be subject of future work.

The perpendicular mean free paths for the different turbulence models are visualized in Fig. \ref{LperpMH}. Again we obtain similar results
for all considered models. In all cases the perpendicular mean free path increases with rigidity if the latter parameter is small. For high
rigidities the perpendicular mean free paths become rigidity independent as already predicted analytically in \textit{Shalchi} [2014] and
\textit{Shalchi} [2015]. In this case the perpendicular diffusion coefficient approaches asymptotically the so-called {\it Field Line Random
Walk (FLRW) limit}. We like to emphasize that $\delta B_z \neq 0$ for isotropic and Goldreich-Sridhar turbulence whereas $\delta B_z = 0$
for the other models. This could cause a difference in the perpendicular diffusion coefficient.

In Fig. \ref{RatioMH} we show the ratio $\lambda_{\perp} / \lambda_{\parallel}$ for the different turbulence models. For low rigidities
the ratio $\lambda_{\perp} / \lambda_{\parallel}$ is approximately constant as expected (see again \textit{Shalchi} [2014] and
\textit{Shalchi} [2015]). Rigidity dependence and magnitude of the perpendicular mean free path depend only weakly on the
chosen turbulence model. An exception is the NRMHD model where the parallel diffusion coefficient is different compared to
the other models. A possible explanation for that difference is provided above.

A direct relation between parallel and perpendicular mean free paths is predicted by the UNLT theory of \textit{Shalchi} [2010] where
the rigidity does not explicitly enter the corresponding integral equation. Therefore, we show the perpendicular mean free path versus
the parallel mean free path for the different turbulence models in Fig. \ref{LperpMHvsLpara}. As predicted by the aforementioned theory,
we find $\lambda_{\perp} \sim \lambda_{\parallel}$ for short parallel mean free path and $\lambda_{\perp} = const$ for long parallel mean
free path.

So far we did not discuss the noisy slab model. The reason is that this model provides very different results if it comes to the perpendicular
diffusion coefficient. This difference was already predicted analytically by the UNLT theory (see \textit{Shalchi} [2015]). The latter theory
states that in the limit $\lambda_{\parallel}/l_{\parallel} \rightarrow 0$ and for small Kubo numbers, the ratio of the two mean free paths
is given by
\begin{linenomath}
\begin{equation}
\frac{\lambda_{\perp}}{\lambda_{\parallel}} = \left[ \frac{\pi}{2} C(s) a^2 \frac{l_{\parallel}}{l_{\perp}} \frac{\delta B^2}{B_0^2} \right]^2
\label{noisyslabana1}
\end{equation}
\end{linenomath}
for the noisy slab model. In the current paper we set $l_{\parallel}/l_{\perp} = 0.5$, $\delta B^2 / B_0^2 = 1$, $a^2 = 1$, and $s=5/3$
leading to $C (s=5/3) \approx 0.12$. In this case Eq. (\ref{noisyslabana1}) provides $\lambda_{\perp} / \lambda_{\parallel} \approx 0.9 \times 10^{-2}$.
For the limit $\lambda_{\parallel}/l_{\parallel} \rightarrow \infty$ and small Kubo numbers we expect to find the quasilinear scaling (see again
\textit{Shalchi} [2015]) and the perpendicular mean free path is given by
\begin{linenomath}
\begin{equation}
\frac{\lambda_{\perp}}{l_{\parallel}} = \frac{3 \pi}{2} C(s) a^2 \frac{\delta B^2}{B_0^2}
\label{noisyslabana2}
\end{equation}
\end{linenomath}
for noisy slab turbulence. For the parameter values used in the current paper (see above), this becomes $\lambda_{\perp} / l_{\parallel} \approx 0.57$.
In Figs. \ref{Noisyslab1} and \ref{Noisyslab2} we show these two analytical limits together with the simulations. We can see that the analytical
results are perfectly in agreement with the simulations. According to Table \ref{MHnoisyslab} we find that the parallel mean free path is
similar compared to the other turbulence models. The ratio $\lambda_{\perp} / \lambda_{\parallel}$, however, is much smaller for noisy slab
turbulence as predicted and explained in \textit{Shalchi} [2015].
\section{Testing our Results by using the PADIAN Code}
To check the validity and accuracy of the results presented in the previous section, we perform test-particle simulations also by
using a different code. In \textit{Tautz} [2010] the so-called PADIAN code was developed. This code is similar compared to
the code described above. A major difference between the two codes is that for three-dimensional turbulence the PADIAN code still uses
a single sum whereas a double sum is used in the Hussein \& Shalchi code. A single sum can be realized by first determining the
two-dimensional surface element of the unit sphere. This is obtained from the distances between the randomly chosen angles in
Eq. (\ref{polarization}). By sorting the combinations into a matrix with increasing values in the rows and columns, the distances
between neighboring numbers---and thus the two-dimensional volume elements---can be calculated. While this approach certainly does
not "fill" the surface with values, it is nevertheless a viable alternative to the double sum introduced in Eq. (\ref{amp}).

\begin{figure}
\centering 
\includegraphics[width=0.48\textwidth]{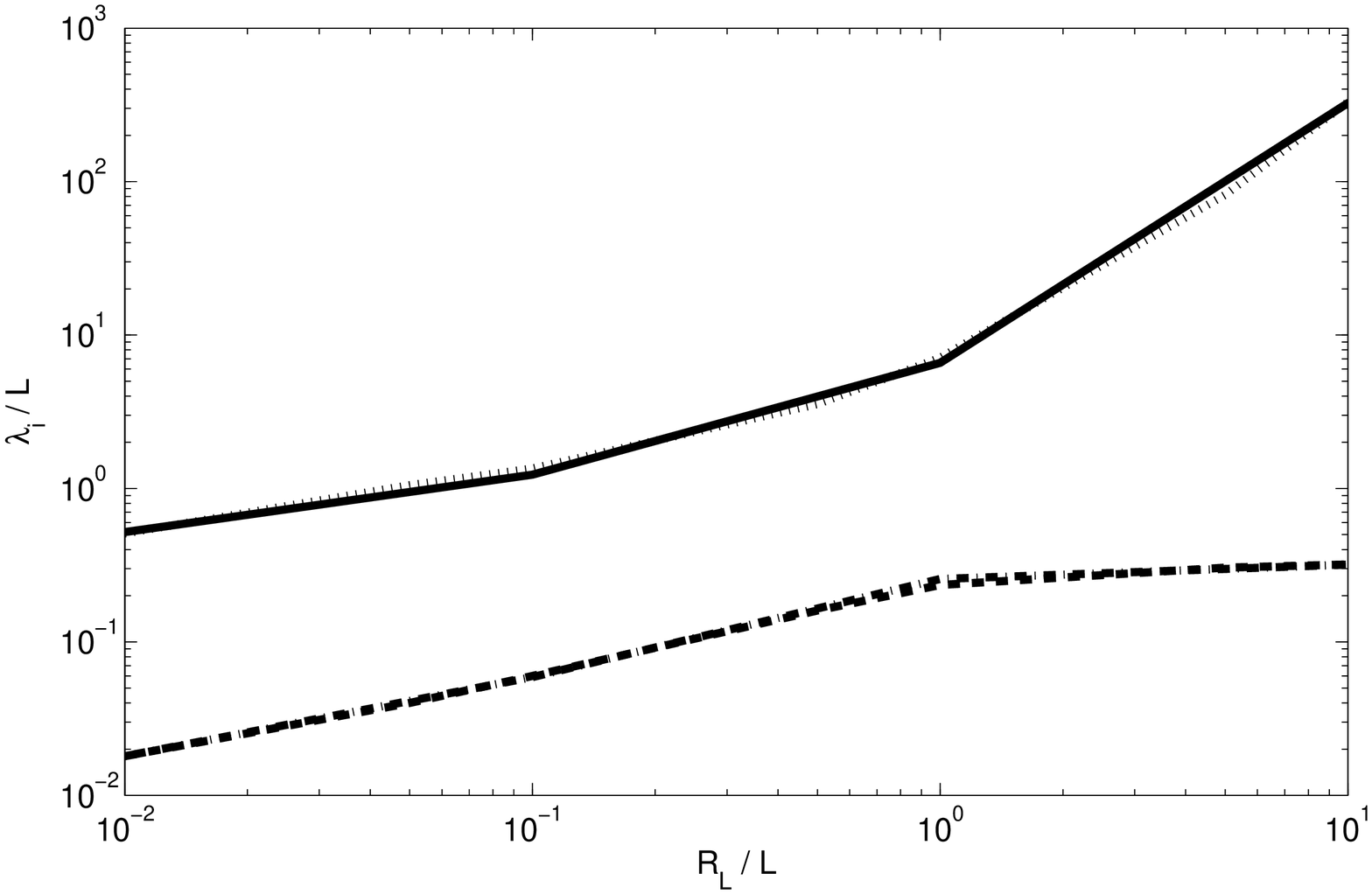}
\caption{The mean free paths for slab/2D turbulence. Shown are the parallel mean free paths from the Hussein \& Shalchi
simulations (dotted line) and PADIAN simulations (solid line) as well as the perpendicular mean free paths from the Hussein \& Shalchi
simulations (dashed line) and PADIAN simulations (dash-dotted line).}
\label{Comparecomp}
\end{figure}

\begin{figure}
\centering 
\includegraphics[width=0.48\textwidth]{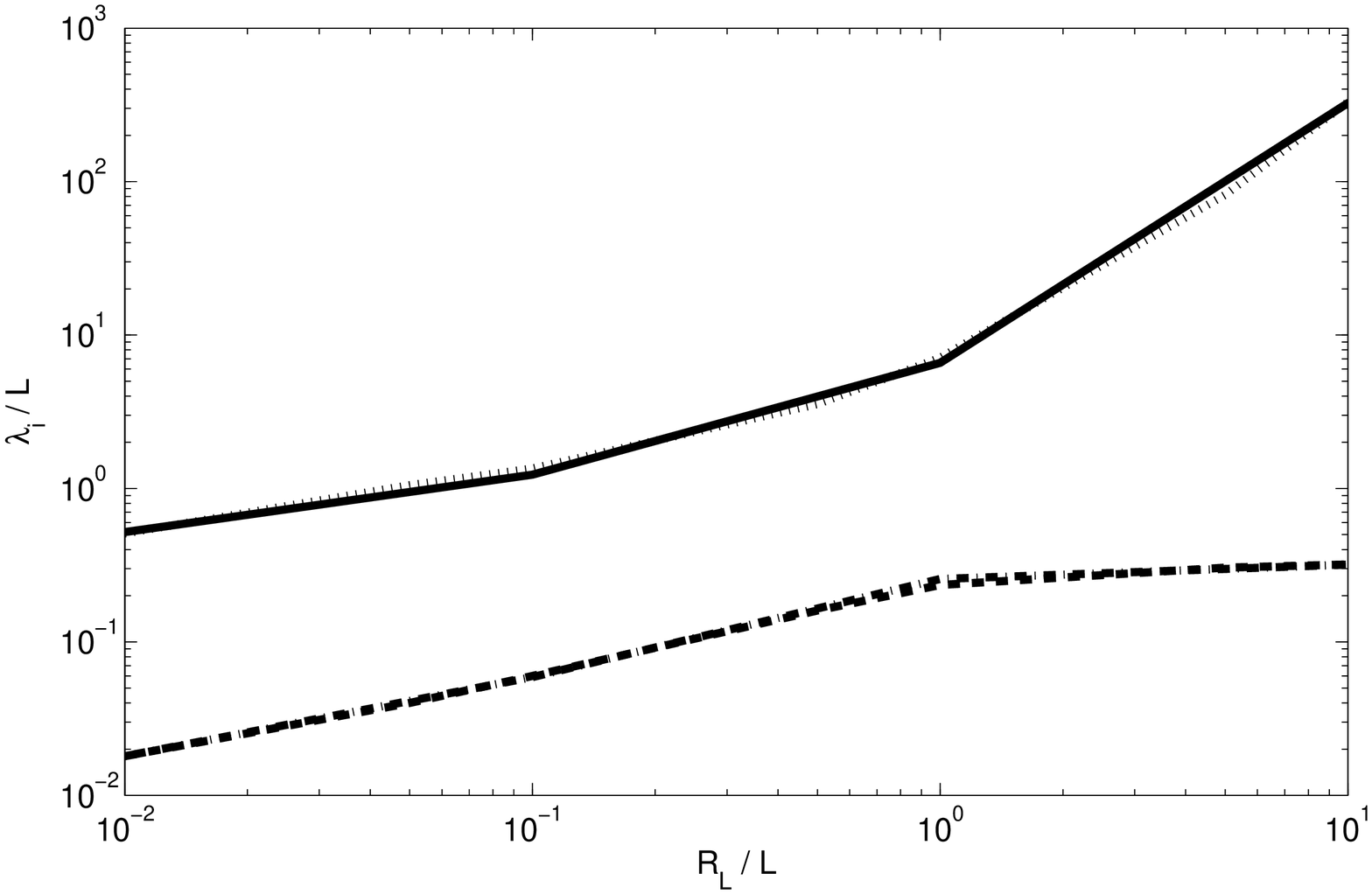}
\caption{The mean free paths for isotropic turbulence. Shown are the parallel mean free paths from the Hussein \& Shalchi
simulations (dotted line) and PADIAN simulations (solid line) as well as the perpendicular mean free paths from the Hussein \& Shalchi
simulations (dashed line) and PADIAN simulations (dash-dotted line).}
\label{Compareiso}
\end{figure}

\begin{figure}
\centering 
\includegraphics[width=0.48\textwidth]{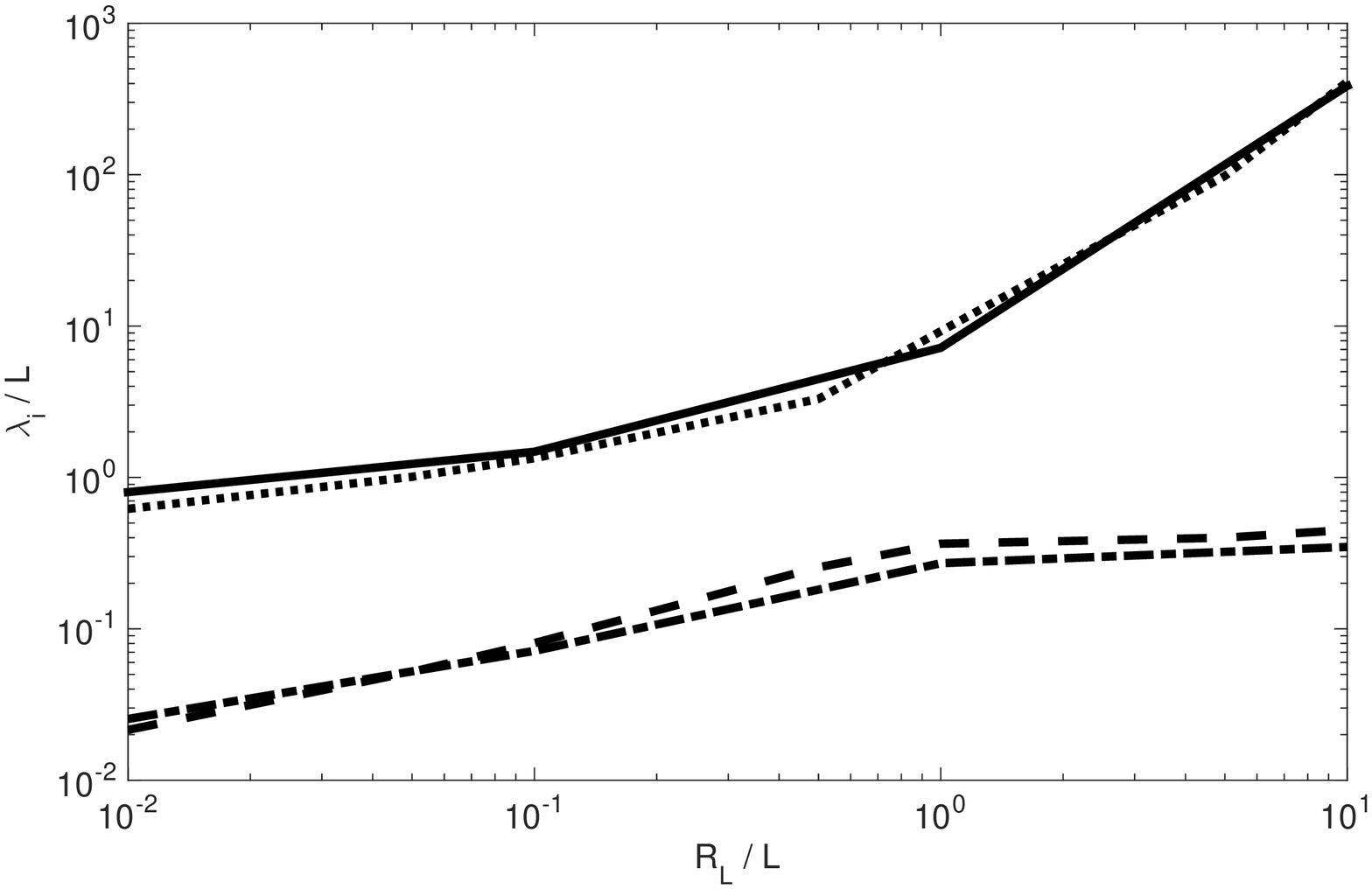}
\caption{The mean free paths for Goldreich-Sridhar turbulence. Shown are the parallel mean free paths from the Hussein \& Shalchi
simulations (dotted line) and PADIAN simulations (solid line) as well as the perpendicular mean free paths from the Hussein \& Shalchi
simulations (dashed line) and PADIAN simulations (dash-dotted line).}
\label{Comparegold}
\end{figure}

\begin{figure}
\centering 
\includegraphics[width=0.48\textwidth]{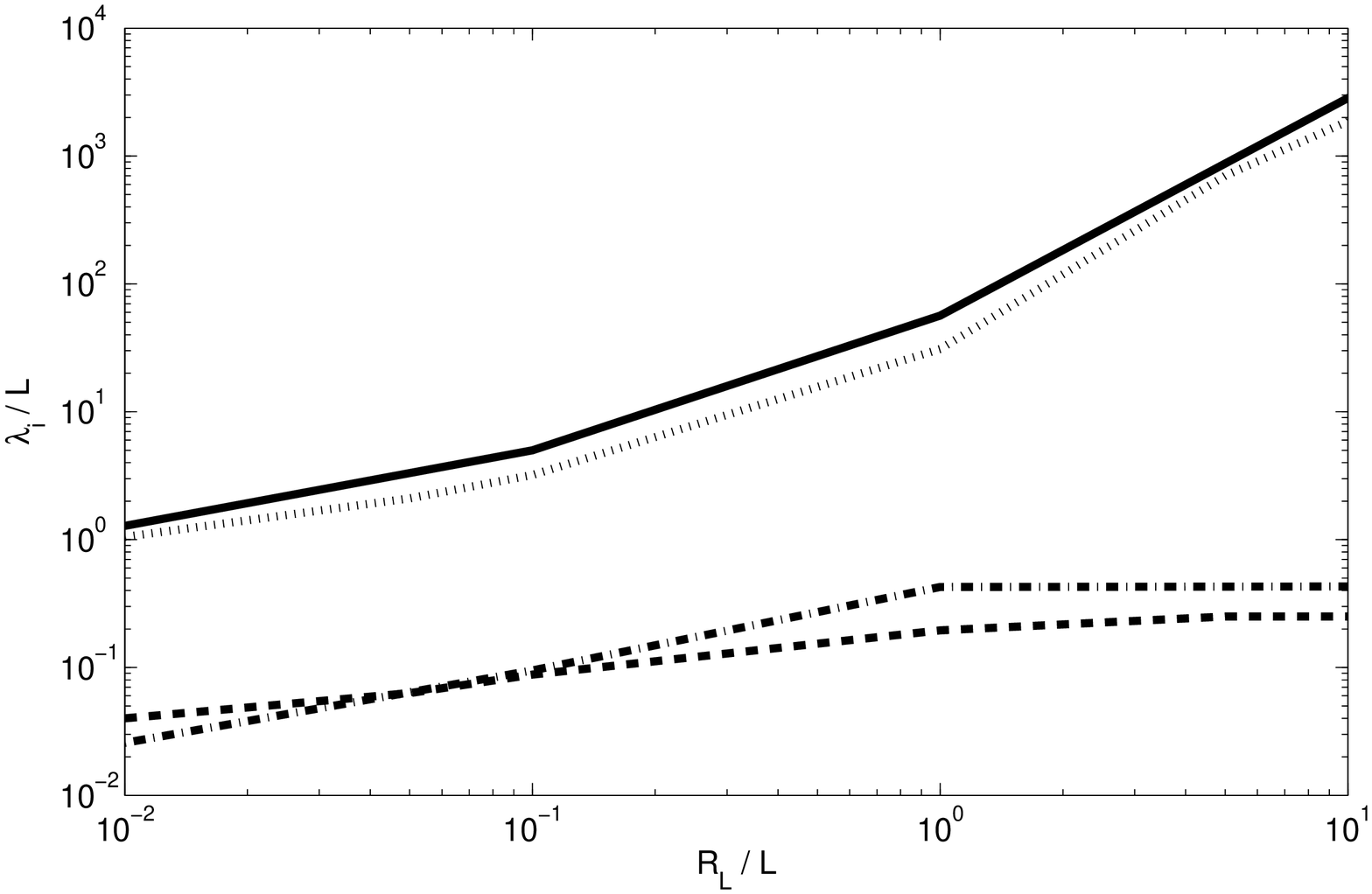}
\caption{The mean free paths for NRMHD turbulence. Shown are the parallel mean free paths from the Hussein \& Shalchi
simulations (dotted line) and PADIAN simulations (solid line) as well as the perpendicular mean free paths from the Hussein \& Shalchi
simulations (dashed line) and PADIAN simulations (dash-dotted line).}
\label{Comparenoisy}
\end{figure}

Again we compute the different diffusion coefficients and compare the numerical findings with the results obtained above. In Tables
\ref{MHcomposite}-\ref{MHnoisy} we summarize the PADIAN results for the individual turbulence models. In Fig. \ref{Comparecomp} we
compare the parallel and perpendicular mean free paths obtained from the two different simulations for two component turbulence.
Obviously we find an almost perfect agreement confirming the validity of our numerical work. In Fig. \ref{Compareiso} the same comparison
is shown but for isotropic turbulence. Again the agreement obtained by using the two different numerical tools is almost perfect.
For anisotropic three-dimensional models such as Goldreich-Sridhar turbulence or NRMHD turbulence, it is more difficult to perform
test-particle simulations and, in this case, there is a major technical difference between the two codes. As described above, the
Hussein \& Shalchi code uses a double sum whereas the PADIAN code is still using a single sum as in the case of reduced dimensionality.
In Fig. \ref{Comparegold} we compare our results for the Goldreich-Sridhar model and Fig. \ref{Comparenoisy} for the NRMHD model.
We can see that the results are very close although not as close as before. Still the agreement allows us to conclude that our numerical
findings are accurate. Furthermore, it seems to be possible to perform the simulations by using a single instead of a double sum even if
the turbulence is three dimensional.
\section{Summary and Conclusion}
We studied the transport of energetic particles interacting with magnetic turbulence. In order to compute diffusion parameters describing
the transport, one has to specify the properties of the magnetic correlation tensor. In the current paper we calculated spatial diffusion
coefficients for different turbulence models, namely
\begin{itemize}
\item The slab/2D composite model,
\item Isotropic turbulence,
\item A model based on \textit{Goldreich and Sridhar} [1995] scaling,
\item The NRMHD model of \textit{Ruffolo and Matthaeus} [2013],
\item A noisly slab model.
\end{itemize}
For all those models we computed the parallel mean free path $\lambda_{\parallel}$, the perpendicular mean free path $\lambda_{\perp}$,
and the ratio $\lambda_{\perp}/\lambda_{\parallel}$.

\begin{figure}
\centering 
\includegraphics[width=0.48\textwidth]{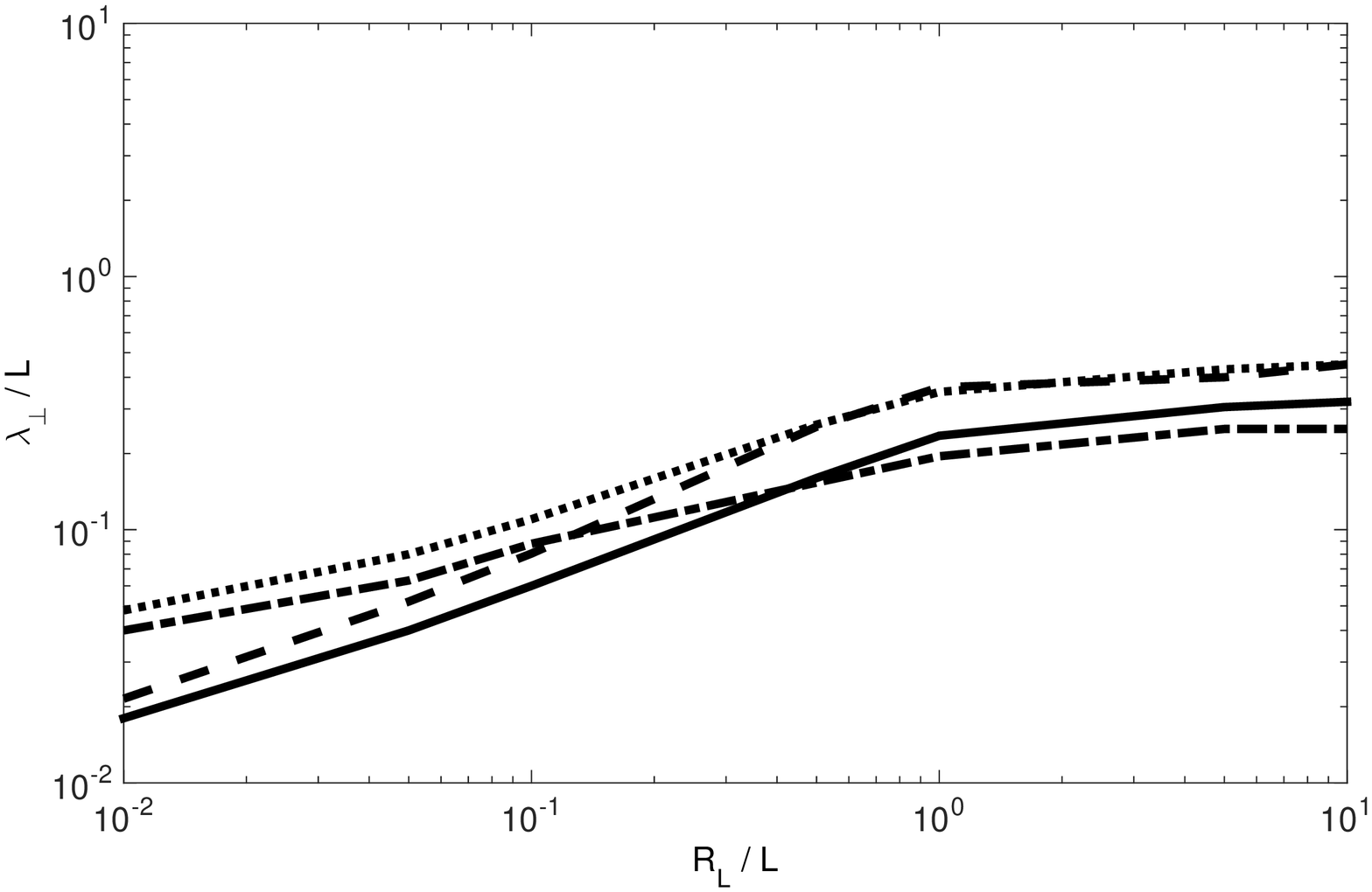}
\caption{The perpendicular mean free paths from the Hussein \& Shalchi simulations. Shown are the results obtained for
slab/2D turbulence (dotted line), isotropic turbulence (solid line), Goldreich-Sridhar turbulence (dashed line), and
NRMHD turbulence (dash-dotted line).}
\label{LperpMH}
\end{figure}

\begin{figure}
\centering 
\includegraphics[width=0.48\textwidth]{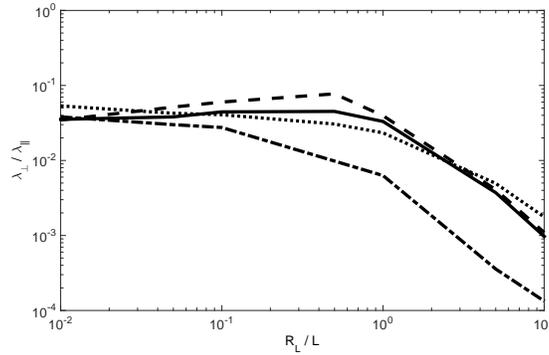}
\caption{The ratio $\lambda_{\perp}/\lambda_{\parallel}$ from the Hussein \& Shalchi simulations. Shown are the results
obtained for slab/2D turbulence (dotted line), isotropic turbulence (solid line), Goldreich-Sridhar turbulence (dashed line),
and NRMHD turbulence (dash-dotted line).}
\label{RatioMH}
\end{figure}

\begin{figure}
\centering 
\includegraphics[width=0.48\textwidth]{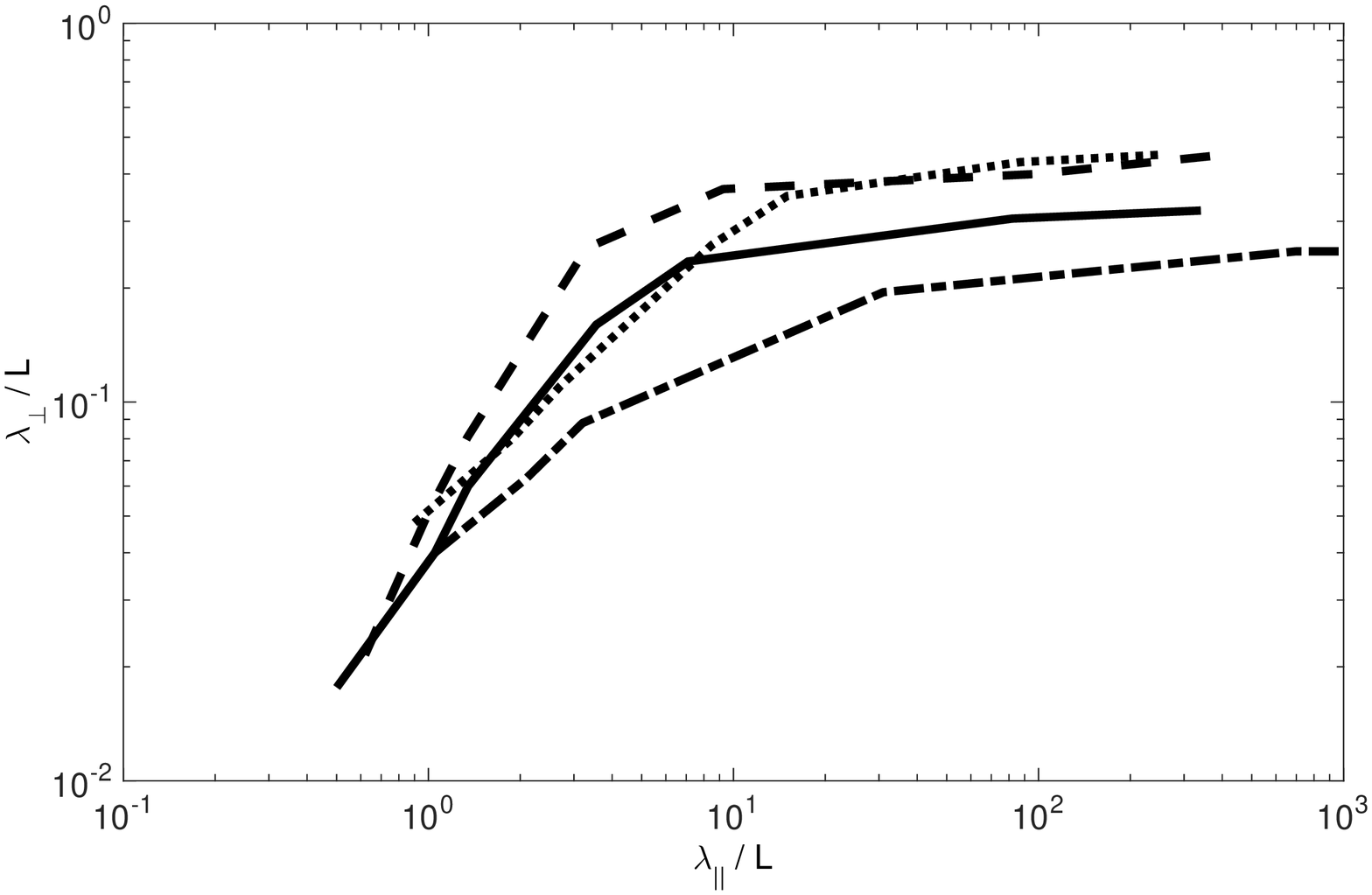}
\caption{The perpendicular mean free paths $\lambda_{\perp}/L$ versus the parallel mean free paths $\lambda_{\parallel}/L$
from the Hussein \& Shalchi simulations. Shown are the results obtained for slab/2D turbulence (dotted line), isotropic
turbulence (solid line), Goldreich-Sridhar turbulence (dashed line), and NRMHD turbulence (dash-dotted line).}
\label{LperpMHvsLpara}
\end{figure}

\begin{figure}
\centering 
\includegraphics[width=0.48\textwidth]{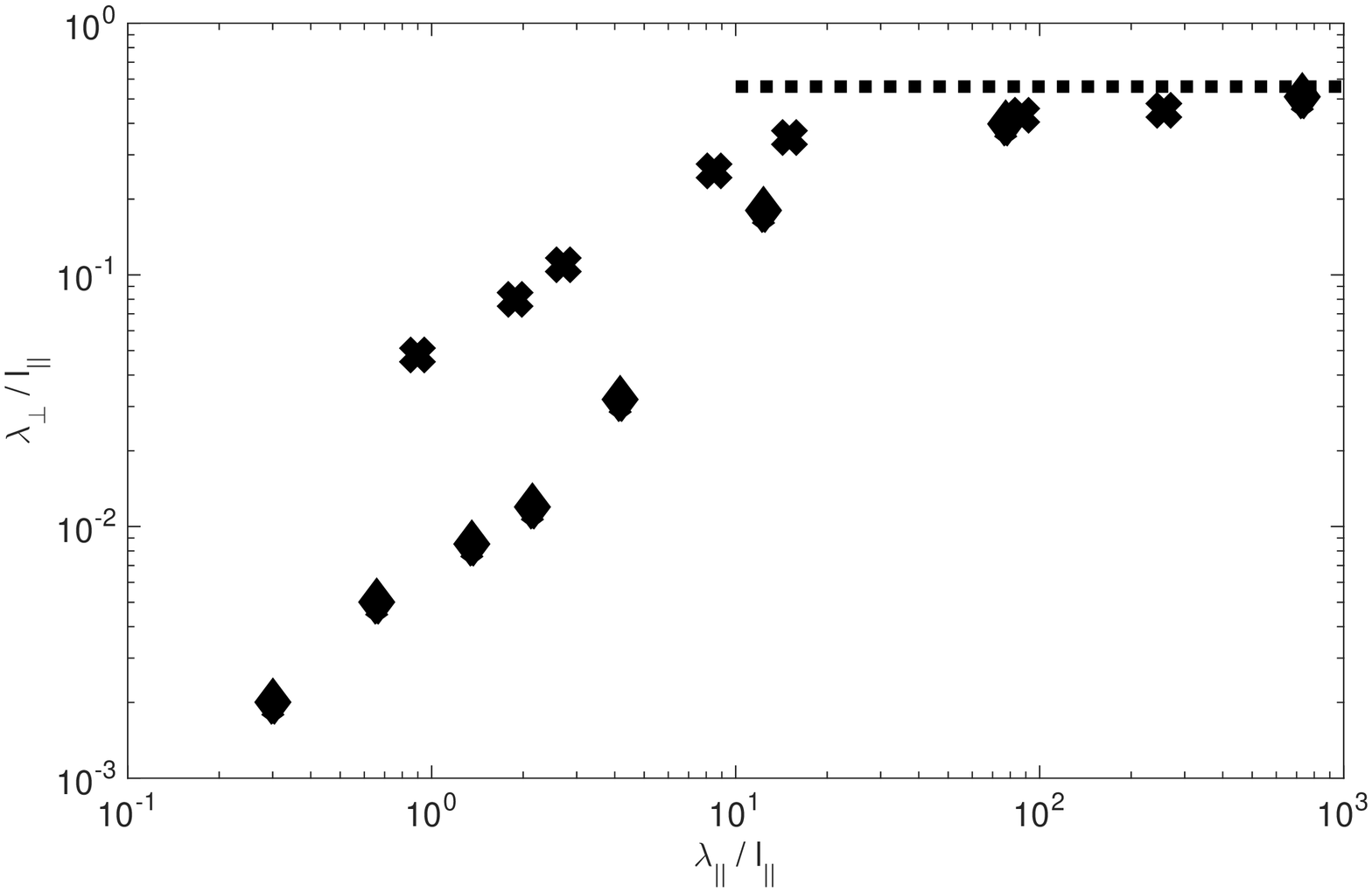}
\caption{The perpendicular mean free paths $\lambda_{\perp}/l_{\parallel}$ versus the parallel mean free paths $\lambda_{\parallel}/l_{\parallel}$
for noisy slab turbulence. Shown are the simulations obtained by employing the Hussein \& Shalchi code (diamonds) and the
analytical results (dotted line) represented by Eq. (\ref{noisyslabana2}). For the sake of comparison we have also shown the
simulations for slab/2D turbulence (crosses).}
\label{Noisyslab1}
\end{figure}

\begin{figure}
\centering 
\includegraphics[width=0.48\textwidth]{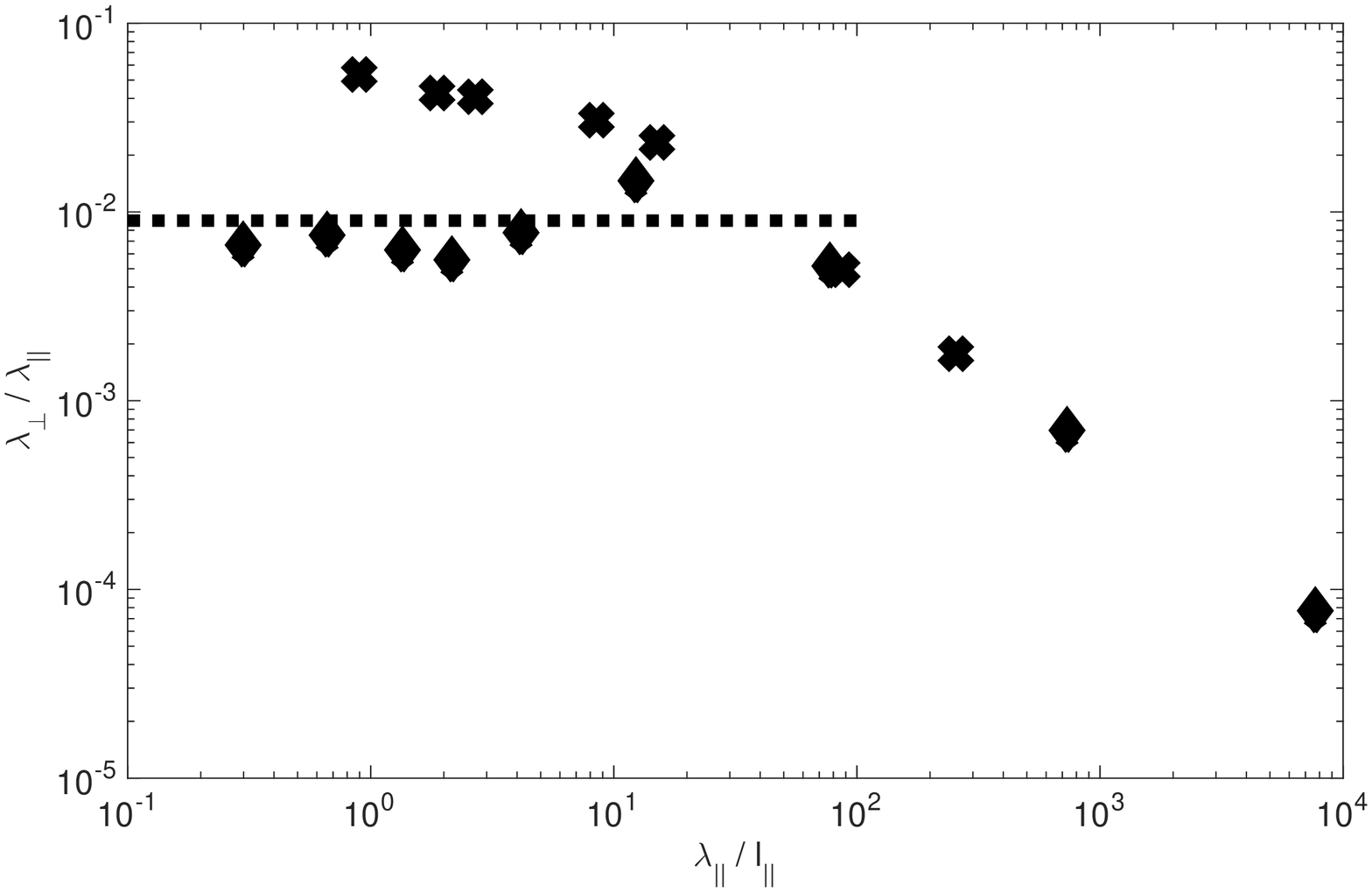}
\caption{The ratio $\lambda_{\perp}/\lambda_{\parallel}$ versus the parallel mean free paths $\lambda_{\parallel}/l_{\parallel}$
for noisy slab turbulence. Shown are the simulations obtained by employing the Hussein \& Shalchi code (diamonds) and the
analytical results (dotted line) represented by Eq. (\ref{noisyslabana1}). For the sake of comparison we have also shown the
simulations for slab/2D turbulence (crosses).}
\label{Noisyslab2}
\end{figure}

We have shown that for all considered turbulence models, the diffusion coefficients have a similar rigidity dependence
and only the absolute values of the diffusion coefficients are different. This conclusion is in agreement with the analytical findings
obtained in \textit{Shalchi} [2014] and \textit{Shalchi} [2015] based on the UNLT theory. The latter theory predicts that the perpendicular
diffusion coefficient is directly proportional to the parallel diffusion coefficient for small rigidities and becomes rigidity independent
for higher rigidities. This is exactly what we found numerically in the current paper (see Figs. \ref{LperpMH}-\ref{Noisyslab2}).
According to our present work, the influence of detailed turbulence properties on the rigidity dependence is minor.
As already proposed in \textit{Shalchi} [2014] and \textit{Shalchi} [2015], only fundamental properties of turbulence such as the
length scales and magnetic fields control the diffusion coefficients. As argued in \textit{Shalchi} [2015], the important
quantity controlling the transport is the Kubo number. If the latter number is extreme, the diffusion parameters are extreme
as well. This is in particular the case for models with reduced dimensionality such as the slab or the two-dimensional model.
Of course, the statements made in the current paper do not apply for such turbulence models.

In the current paper we have only explored particle transport for certain parameter regimes (e.g., we have assumed that
$\delta B = B_0$). Analytical theories (see again \textit{Shalchi} [2015]) predict that perpendicular diffusion does only
depends on the parallel diffusion coefficient and the Kubo number. The latter number is directly proportional to the ratio
$\delta B / B_0$. Therefore, we expect that changing the magnetic field ratio will have a strong influence on the magnitudes
of the two diffusion parameters. However, qualitatively the diffusion parameters should be similar for weak turbulence
(e.g., for $\delta B/B_0 = 0.1$ or $\delta B/B_0 = 0.3$). The latter statement is supported by the numerical work presented
in \textit{Hussein and Shalchi} [2014a] where the influence of the magnetic field ratio on the transport had been explored.

To explore the influence of turbulence on the diffusion coefficients of energetic particles was also subject of
previous work (see, e.g., \textit{Giacalone and Jokipii} [1999], \textit{Sun and Jokipii} [2011]). The numerical results
presented in this previous work is similar compared to our findings. However, we have added more turbulence models
(e.g., the NRMHD model) and we have used a more appropriate spectrum at large scales to ensure that all fundamental
scales of turbulence (e.g., the ultra-scale) are finite.

For the first time we have explored numerically transport of particles in noisy slab turbulence. For the particular turbulence
model we found a very small ratio $\lambda_{\perp}/\lambda_{\parallel}$ in perfect agreement with the results obtained in
\textit{Shalchi} [2015]. Once more the UNLT theory is confirmed by numerical work.

To double-check our findings and to estimate how accurate our numerical results are, we have employed a second test-particle code, namely
the so-called PADIAN code originally developed in \textit{Tautz} [2010]. We have shown that this second code provides similar results.
Only small variations can be found if the results provided by the two codes are compared with each other. Therefore, we conclude that
our numerical findings are correct and accurate enough to draw conclusions concerning the influence of turbulence on the transport of
energetic particles.
\begin{acknowledgements}
M. Hussein and A. Shalchi acknowledge support by the Natural Sciences and Engineering Research Council (NSERC)
of Canada and national computational facility provided by WestGrid. We are also grateful to S. Safi-Harb for
providing her CFI-funded computational facilities for code tests and for some of the simulation runs presented here.
\end{acknowledgements}
{}

\end{article}

\begin{table}
\caption{The values used in the simulations for slab, two-dimensional, isotropic, NRMHD, Goldreich-Sridhar, and noisy slab turbulence.}
\begin{center}
\begin{tabular}{llllll}
\hline
Turbulence model	& $\eta_n$	& $\alpha_n$ 	& $\Phi_n$ 	& Wave numbers												& Energy range spectral index	\\
\hline
Slab				& $1$		& $0$	 		& Random 	& $k_n=l_{slab} k_{\parallel}$ 								& $0$							\\
Two-dimensional		& $0$		& $0$	 		& Random 	& $k_n=l_{2D} k_{\perp}$ 									& $2$							\\
Isotropic			& Random	& Random 		& Random 	& $k_n=l_0 k$ 												& $3$							\\
\hline
NRMHD				& $0$		& $0$	 		& Random 	& $k_n=l_{\perp}k_{\perp}$,$k_m=l_{\parallel}k_{\parallel}$	& $3$ 							\\
Goldreich-Sridhar	& $0$		& Random 		& Random 	& $k_n=l k_{\perp}$,$k_m=l k_{\parallel}$ 					& $2$							\\
Noisy slab model	& $0$		& $0$			& Random 	& $k_n=l_{\perp}k_{\perp}$,$k_m=l_{\parallel}k_{\parallel}$	& $0$							\\
\hline
\end{tabular}
\end{center}
\label{anglestab}
\end{table}

\begin{table}
\caption{Test particle simulations for slab/2D turbulence. For the energy range spectral index of the two-dimensional modes we used $q=2$.
The two scales are assumed to be equal $l_{2D}=l_{slab}$ and we set $\delta B_{slab}^2 / B_0^2 = 0.2$ and $\delta B_{2D}^2 / B_0^2 = 0.8$
as suggested by Bieber et al. (1996). Listed are the results obtained by using the Hussein \& Shalchi code (HS) and the PADIAN code (P).}
\begin{center}
\begin{tabular}{llllllll}
\hline
$R_L / l_{slab}$						& $0.01$	& $0.05$ 	& $0.1$ 	& $0.5$ 	& $1.0$		& $5.0$		& $10.0$	\\
\hline
$\lambda_{\parallel}^{HS} / l_{slab}$	& $0.90$	& $1.88$ 	& $2.7$ 	& $8.5$ 	& $15.0$	& $87.0$	& $255$		\\
$\lambda_{\perp}^{HS} / l_{slab}$		& $0.048$	& $0.08$ 	& $0.11$ 	& $0.26$ 	& $0.35$	& $0.43$	& $ 0.45$	\\
\hline
$\lambda_{\parallel}^{P} / l_{slab}$	& $0.932$	& $-$		& $2.581$ 	& $-$		& $15.0$ 	& $-$		& $266$ 	\\
$\lambda_{\perp}^{P} / l_{slab}$		& $0.049$	& $-$		& $0.123$ 	& $-$		& $0.384$ 	& $-$		& $0.456$ 	\\
\hline
\end{tabular}
\end{center}
\label{MHcomposite}
\end{table}

\begin{table}
\caption{Test particle simulations for isotropic turbulence. For the energy range spectral index we used $q=3$ as explained in the text.
Listed are the results obtained by using the Hussein \& Shalchi code (HS) and the PADIAN code (P).}
\begin{center}
\begin{tabular}{llllllll}
\hline
$R_L / l_{0}$							& $0.01$	& $0.05$ 	& $0.1$ 	& $0.5$ 	& $1.0$		& $5.0$		& $10.0$	\\
\hline
$\lambda_{\parallel}^{HS} / l_{0}$		& $0.51$	& $1.05$ 	& $1.35$ 	& $3.55$ 	& $7.1$		& $82.0$	& $330$		\\
$\lambda_{\perp}^{HS} / l_{0}$			& $0.018$	& $0.04$ 	& $0.06$ 	& $0.16$ 	& $0.235$	& $0.305$	& $0.32$	\\
\hline
$\lambda_{\parallel}^{P} / l_{0}$		& $0.52$	& $-$		& $1.23$ 	& $-$		& $6.58$ 	& $-$		& $325$ 	\\
$\lambda_{\perp}^{P} / l_{0}$			& $0.018$	& $-$		& $0.059$ 	& $-$		& $0.257$ 	& $-$		& $0.319$ 	\\
\hline
\end{tabular}
\end{center}
\label{MHisotropic}
\end{table}

\begin{table}
\caption{Test particle simulations for Goldreich-Sridhar turbulence. For the energy range spectral index we used $q=2$.
Listed are the results obtained by using the Hussein \& Shalchi code (HS) and the PADIAN code (P).}
\begin{center}
\begin{tabular}{llllllll}
\hline
$R_L / l$						& $0.01$	& $0.05$	& $0.1$ 	& $0.5$ 	& $1.0$ 	& $5.0$		& $10.0$	\\
\hline
$\lambda_{\parallel}^{HS} / l$	& $0.62$	& $1.01$	& $1.34$ 	& $3.3$ 	& $9.25$ 	& $98.0$	& $410.0$	\\
$\lambda_{\perp}^{HS} / l$		& $0.0215$	& $0.052$ 	& $0.0805$ 	& $0.255$ 	& $0.365$	& $0.40$	& $0.45$	\\
\hline
$\lambda_{\parallel}^{P} / l$	& $0.800$	& $-$		& $1.478$	& $-$		& $7.186$ 	& $-$		& $386$ 	\\
$\lambda_{\perp}^{P} / l$		& $0.0251$	& $-$		& $0.0715$ 	& $-$		& $0.2711$ 	& $-$		& $0.348$ 	\\
\hline
\end{tabular}
\end{center}
\label{MHgoldreich2}
\end{table}

\begin{table}
\caption{Test particle simulations for NRMHD turbulence. For the energy range spectral index we used $q=3$ and we assumed that $l_{\parallel}=l_{\perp}$.
Listed are the results obtained by using the Hussein \& Shalchi code (HS) and the PADIAN code (P).}
\begin{center}
\begin{tabular}{lllllll}
\hline
$R_L / l_{\perp}$						& $0.01$ 	& $0.05$ 	& $0.1$ 	& $1.0$		& $5.0$		& $10$		\\
\hline
$\lambda_{\parallel} / l_{\perp}$		& $1.05$ 	& $2.1$ 	& $3.2$ 	& $31$		& $700$		& $1875$	\\
$\lambda_{\perp} / l_{\perp}$			& $0.04$ 	& $0.063$ 	& $0.088$ 	& $0.195$	& $0.25$	& $0.25$	\\
\hline
$\lambda_{\parallel} / l_{\perp}$		& $1.279$	& $-$		& $5.002$	& $56.47$	& $-$		& $2843$	\\
$\lambda_{\perp} / l_{\perp}$			& $0.0257$	& $-$		& $0.0948$ 	& $0.4255$ 	& $-$		& $0.4302$ 	\\
\hline
\end{tabular}
\end{center}
\label{MHnoisy}
\end{table}

\begin{table}
\caption{The Hussein \& Shalchi simulations for noisy slab turbulence. For the scale ratio we have assumed $l_{\parallel}/l_{\perp} = 0.5$
and the used spectrum corresponds to the one used for the standard slab model.}
\begin{center}
\begin{tabular}{llllllllll}
\hline
$R_L / l_{\parallel}$						& $0.001$ 	& $0.01$ 	& $0.1$ 	& $0.316$	& $1.0$		& $3.16$	& $10.0$	& $31.6$	& $100$		\\
\hline
$\lambda_{\parallel} / l_{\parallel}$		& $0.3$ 	& $0.66$ 	& $1.35$ 	& $2.15$	& $4.15$	& $12.4$	& $77$		& $730$		& $7700$	\\
$\lambda_{\perp} / l_{\parallel}$			& $0.002$ 	& $0.005$ 	& $0.0085$ 	& $0.012$	& $0.032$	& $0.181$	& $0.40$	& $0.51$	& $0.59$	\\
\hline
\end{tabular}
\end{center}
\label{MHnoisyslab}
\end{table}

\end{document}